# Cross-Attention Multimodal Fusion for Breast Cancer Diagnosis: Integrating Mammography and Clinical Data with Explainability

Muhaisin Tiyumba Nantogmah
*Department of Computer Science*
*University for Development Studies*
Northern Region, Ghana.
muhaisintiyumbana1848@uds.edu.gh

Prof Abdul-Barik Alhassan
*Department of Computer Science*
*University for Development Studies*
Northern Region, Ghana.
abarik@uds.edu.gh

Dr Salamudeen Alhassan
*Department of Computer Science*
*University for Development Studies*
Northern Region, Ghana.
salamudeen.alhassan@uds.edu.gh

*Abstract*–A precise assessment of the risk of breast lesions can greatly lower it and assist physicians in choosing the best course of action. To categorise breast lesions, the majority of current computer-aided systems only use characteristics from mammograms. Although this method is practical, it does not completely utilise clinical reports' valuable information to attain the best results. When compared to utilising mammography alone, will clinical features greatly enhance the categorisation of breast lesions? How may clinical features and mammograms be combined most effectively? In what ways may explainable AI approaches improve the interpretability and reliability of models used to diagnose breast cancer? To answer these basic problems, a comprehensive investigation is desperately needed. In order to integrate mammography and categorical clinical characteristics, this study examines a number of multimodal deep networks grounded on feature concatenation, co-attention, and cross-attention. The model achieved an AUC-ROC of 0.98, accuracy of 0.96, F1-score of 0.94, precision of 0.92, and recall of 0.95 when tested on publicly accessible datasets (TCGA and CBIS-DDSM).

## I. Introduction

World Health Organization (WHO) presents that breast cancer is classified as the second deadliest cancer among women and causes the most deaths (World Health Organization, 2023). Breast cancer manifests through a variation of symptoms. Identifying specific breast cells that have undergone a malignant transformation is the basis for determining the type of breast cancer. The natural stage of aging leads to genetic deficiency and accounts for 90% of breast cancer patients. The remaining 5-10% of cases are genetic traits that people acquire from their parents through heredity (National Cancer Institute, 2023). Conventional diagnostic techniques for instance mammography, and ultrasound are essentially one-way, but gaps in sensitivity and specificity often lead to false positives or missed diagnoses (Verrus et al., 2023). Additionally, these methods do not use complementary knowledge provided by non-imaging data, such as genetic markers and clinical parameters that are important for understanding the disease as a whole (Dorsey Vélez & King, 2017).

Deep learning (DL) and machine learning (ML) have seen a lot of study and development recently in a variety of industries, including travel (Sahara, Bhave et al., 2020), healthcare (Krizhevsky et al., 2017) and smart parking systems (Sahara, Bawah et al., 2020). They show considerable latent in assisting radiologists to provide further precise diagnoses and lower the performance variance (Freer & Ulissey, 2001). The overall issue of determining whether a lesion is benign or malignant is the main focus of prior research (Shen et al., 2019; Ribli et al., 2018). More focused classification issues, such as the categorisation of lesion forms (mass or calcification) and BI-RADS density (level 1 to 4) have recently been the focus of a number of breast cancer articles (Shen et al., 2019; Matthews et al., 2020; Chougrad et al., 2020). A lesion picture is classified into five classifications in the Shen et al., (2019) study, benign mass, malignant mass, benign calcification, malignant calcification, or background. Lesions in the breasts that occupy three dimensions are called masses. Calcium salt deposits in the breast are known as calcifications. Classifying the kind of lesion initially can aid in the diagnosis of cancer since different forms of lesions (calcification, tumour, etc.) have different characteristics. Another crucial element in the assessment of pathology is breast density (Kerlikowske et al., 2010). This sub-problem was the focus of Matthews et al. (2020), who demonstrated encouraging outcomes. In order to take advantage of any potential association between the labels, Chougrad et al. (2020) integrated data on lesion kinds, breast mass, and pathology in a multilabel environment.

While machine learning and deep learning techniques have done exemplary work in detecting and classifying breast cancer, they have severely hampered the clinical transition. These models "black box" status makes it difficult to understand the root causes of decisions and undermines the trust of healthcare professionals who need transparent and translational outcomes to support evidence-based care (Arrieta et al., 2020; Topol, 2019). As a result, this hinders the widespread use of artificial intelligence in healthcare, especially in critical applications where diagnostic errors can affect life choices. Addressing the need for different ways to create models that are easy for people to understand (Dosi Vélez & Kim, 2017), explainable artificial intelligence (XAI) unravels the mystery of models. Explainable artificial intelligence (XAI) is defined as: (Van Lent et al., 2004) *"How well a system can describe the actions of AI-powered objects in a simulation game."* According to the Defence Advanced Research Projects Agency (DARPA), XAI is an effort to create models that are more understandable and highly accurate for human users to understand, in order to promote trust, understanding, and efficient management of future AI partners.

As far as we are aware, the majority of methods just input mammograms data for pathology categorisation. In this study, we propose an explainable multimodal model which helps improve the pathology classification performance by leveraging multimodal data and provide interpretability for the model diagnosis using the accessible labels of clinical features that correspond with their agreeing mammogram in the CBIS-DDSM catalogue (Lee, Gimenez, Hoogi, Miyake, et al., 2017), Additionally, we evaluate our suggested framework in conjunction with either co-attention or cross-attention to address the issue of missing clinical feature data, which frequently occurs in real-world settings.

In particular, this paper's contributions consist of:

1) To enhance the diagnosis of breast cancer pathology, we integrate imaging and clinical aspects to improve the diagnostic process. In this case, an entrenching taken from a pretrained deep learning model will serve as the imaging features. Breast density, mass form, mass borders, calcification type, and calcification distribution are the five parameters we employ for clinical purposes. These features will be shown as a vector that is created by concatenating five one-hot vectors, each of which represents one of the five features (or a zero vector for each absent feature).
2) Assessment of the use of basic concatenation, attention, and cross-attention in the presence of missing data: Due to differences in clinical practice, clinical characteristics are occasionally either absent or insufficient in real-world settings. We add either a basic concatenation or an attention or cross-attention module to our suggested architecture in order to enable our model to be intelligently adaptable to clinical scenarios and readily integrated into any clinical workflow. We also teach students how to deal with clinical aspects that are absent.
3) Integration of Explainable Artificial Intelligence (XAI) techniques for model interpretability and trustworthiness: To enhance transparency and support clinical decision-making, we apply SHAP, LIME, and Grad-CAM to interpret the model's predictions at both the image and feature levels. These methods provide visual and feature-based explanations that highlight critical imaging regions and influential clinical attributes, enabling clinicians to validate the model's reasoning and increasing trust in its diagnostic recommendations.

II. STUDY METHODOLOGY

*A. Multimodal Fusion Network Framework*

*1) Embedding of Mammogram*

Let $x_i^{(m)} \in \mathbb{R}^{H \times W \times 3}$ denote the pre-processed mammogram for patient $i$, resized to $224 \times 224$ pixels. To capture discriminative mammographic features, we employ a ResNet-50 backbone pretrained on ImageNet (He et al., 2016) and fine-tuned on the CBIS-DDSM dataset. The ResNet-50 architecture incorporates residual connections, enabling training of deep networks without degradation in accuracy by mitigating the vanishing gradient problem.

The output of the ResNet-50 global average pooling (GAP) layer yields a $d_e$-dimensional embedding:

$$e_i = ResNet50\left(x_i^{(m)}\right), \quad e_i \in \mathbb{R}^{d_e}, d_e = 2048. \quad (1)$$

To obtain a compact latent representation, we project $e_i$ into a $d$-dimensional space using an affine transformation followed by a nonlinear activation:

$$\tilde{e}_i = F(W_e^T \cdot e_i + b_e), \quad W_e \in \mathbb{R}^{d_e \times d}, b_e \in \mathbb{R}^d, \quad (2)$$

where $F(\cdot)$ is the Rectified Linear Unit (ReLU) activation, $d = 100$ in our default configuration. This compact embedding facilitates downstream fusion by aligning modality dimensions.

*2) Clinical Embedding*

Structured clinical features, including breast density, lesion shape, margins, calcification type, and calcification distribution, are encoded as one-hot vectors with dimensions corresponding to their categorical granularity (breast density $\in \mathbb{R}^4$, mass shape $\in \mathbb{R}^8$,). When features are absent, the associated vectors are set to zero. For example, mass lesions will be represented by two zero vectors as they lack a calcification distribution or kind. The situation with calcification lesions is comparable. Any missing feature will produce one zero vector in practical applications. In conclusion, the total of the dimensions of five feature vectors will yield 36 dimensions for our clinical embedding.

Let $c_i \in \mathbb{R}^{D_c}$ denote the concatenated clinical feature vector for patient $i$, where $D_c = 36$ in our configuration. We apply a similar projection to that used for imaging:

$$\tilde{c}_i = F(W_c^T \cdot c_i + b_c), \quad W_c \in \mathbb{R}^{D_c \times d}, b_c \in \mathbb{R}^d. \quad (3)$$

This ensures both modalities share a common latent dimensionality, enabling symmetric fusion operations.

*3) Multimodal fusion*

Following the acquisition of projected clinical embedding $\tilde{c}_i$ and projected mammography embedding $\tilde{e}_i$, the network performs cross-modal integration via one of several fusion mechanisms.

*Baseline Concatenation (Intermediate Fusion)*

The simplest approach concatenates the two embeddings:

$$\boldsymbol{k}_i = Concat(\tilde{\boldsymbol{e}}_i, \tilde{\boldsymbol{c}}_i) \in \mathbb{R}^{2d} \quad (4)$$

where $\boldsymbol{k}_i$ is the final concatenated embedding. In this work, intermediate fusion was used; we only concatenated two projected embedding vectors, but alternative techniques may be tried to determine whether they enhance the model's performance any more. This vector is then forwarded through dense layers for joint representation learning.

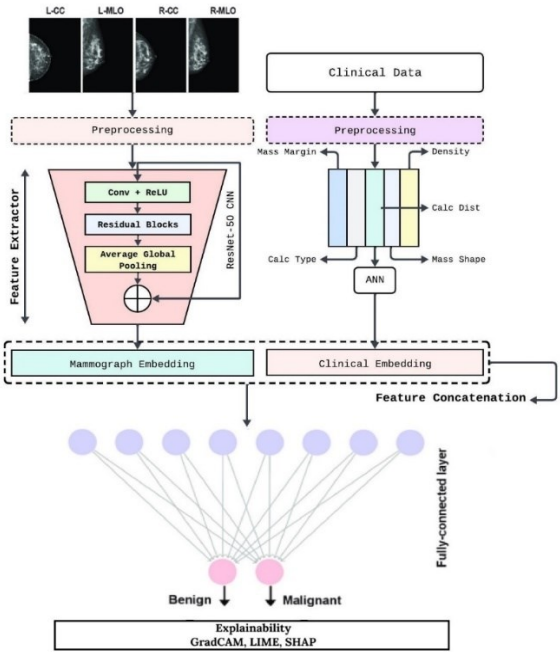

Fig. 1: The general framework of our proposed multimodal fusion model

### B. Co-attention and Cross-attention Strategies

Mammograms are rarely supplied with clinical features in clinical settings, or just a portion of the findings are. Missing value issues result from this. In this paper, we evaluate our model in combination with either the cross-attention module or the co-attention module.

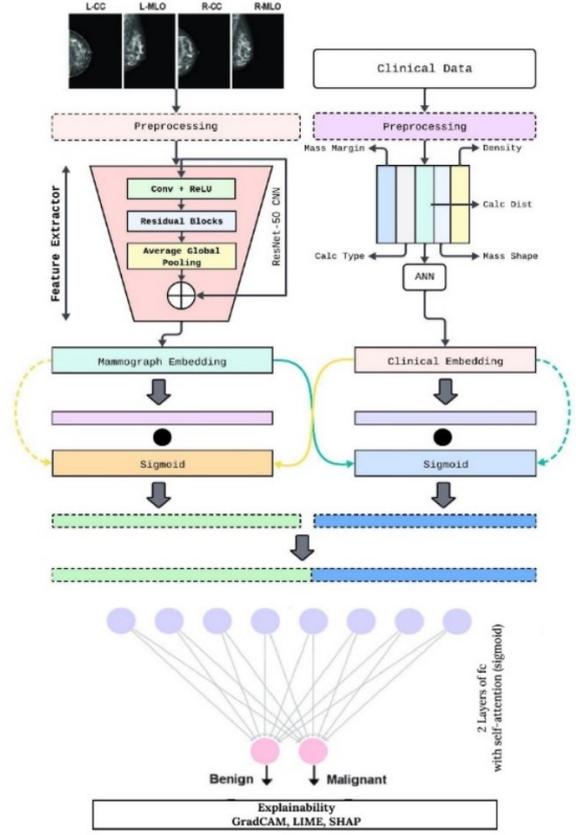

Fig. 2: Co-attention and Cross-attention network design.

We add a cross-attention module or a co-attention module to our suggested structure. Two dashed arrows in the illustration have been eliminated for the cross-attention module.

In order to educate the model, the pertinent details between a mammogram and its clinical aspects, co-attention/cross-attention is utilised in this instance. Our suggested framework in conjunction with the co-attention/cross-attention module is depicted in Figure 2.

*1) Co-attention Approach:*

Co-attention, proposed in (Lu et al., 2019) is a symmetric attention mechanism in which both modalities attend to each other simultaneously. It produces mutually informed representations, where the updated features of each modality are conditioned on the other. This is particularly valuable in our setting because certain radiological patterns (e.g., microcalcifications) only become diagnostically relevant in the context of certain clinical features (e.g., patient age, breast density).

Let:

- $E \in \mathbb{R}^{n_e \times d}$ be the sequence of image features, where $n_e$ is the number of visual tokens (e.g., patches, regions) and ddd the embedding dimension.

- $C \in \mathbb{R}^{n_c \times d}$ be the sequence of clinical feature embeddings, where $n_c$ is the number of discrete clinical tokens.
- $W_Q^e, W_K^e, W_V^e \in \mathbb{R}^{d \times d_k}$ are learnable projection matrices for queries, keys, and values from the image modality.
- $W_Q^c, W_K^c, W_V^c \in \mathbb{R}^{d \times d_k}$ are the analogous matrices for the clinical modality.

We first compute the bidirectional affinity scores between the modalities:

$$S_{ec} = \frac{(EW_Q^e)(CW_K^c)^T}{\sqrt{d_k}} \in \mathbb{R}^{n_e \times n_c}, \quad (5)$$

$$S_{ce} = \frac{(CW_Q^c)(EW_K^e)^T}{\sqrt{d_k}} \in \mathbb{R}^{n_c \times n_e}, \quad (6)$$

Here:

- $S_{ec}[i,j]$ measures how much the $i$-th image token attends to the $j$-th clinical token.
- $S_{ce}[p,q]$ measures how much the $p$-th clinical token attends to the $q$-th image token.

The attention weights are obtained by row-wise softmax normalization:

$$\alpha_{ec} = softmax(S_{ec}) \quad \text{over clinical tokens,}$$

$$\alpha_{ce} = softmax(S_{ce}) \quad \text{over image tokens.}$$

We then compute the contextually updated features:

$$E' = \alpha_{ec}(CW_V^c) \in \mathbb{R}^{n_e \times d_k}, \quad (7)$$

$$C' = \alpha_{ce}(EW_V^e) \in \mathbb{R}^{n_c \times d_k}. \quad (8)$$

Finally, we project back to the original dimension $d$ and apply residual connections and layer normalization:

$$\hat{E} = LN(E + E'W_O^e), \quad \hat{C} = LN(C + C'W_O^c), \quad (9)$$

where $W_O^e, W_O^c \in \mathbb{R}^{d_k \times d}$ are learnable output projections.

The co-attention mechanism ensures that every image token is re-weighted based on *all* clinical tokens, and every clinical token is re-weighted based on *all* image tokens. This symmetric interaction is particularly suited for datasets where both modalities contain complementary diagnostic cues.

*2) Cross-attention Approach*

Cross-attention, proposed in (Abavisani et al., 2020). In contrast, Cross-Attention is asymmetric: one modality acts as the *query* and the other as the *key/value*. This allows targeted conditioning for example, allowing the image features to be selectively enhanced based on only the most relevant clinical features, without reciprocally altering the clinical representations (or vice versa).

Let us define the case where image features are conditioned on clinical features.

We compute the queries from the image modality and keys/values from the clinical modality:

$$Q_E = EW_Q^e \in \mathbb{R}^{n_e \times d_k}, \quad (10)$$

$$K_C = CW_K^c \in \mathbb{R}^{n_c \times d_k}, \quad (11)$$

$$V_C = CW_V^e \in \mathbb{R}^{n_c \times d_k}.$$

The attention score matrix is:

$$S_{E \leftarrow C} = \frac{Q_E K_C^T}{\sqrt{d_k}} \in \mathbb{R}^{n_e \times n_c}. \quad (12)$$

We normalize across clinical tokens:

$$\alpha_{E \leftarrow C} = softmax(S_{E \leftarrow C}), \quad (13)$$

and compute the updated image features:

$$E' = \alpha_{E \leftarrow C} V_C. \quad (14)$$

The updated image embedding is:

$$\hat{E} = LN(E + E'W_O^e). \quad (15)$$

Reverse conditioning (clinical on image) is achieved by swapping the roles of EEE and CCC in equations (12) – (15).

Cross-attention is more directional, allowing us to enforce a hierarchy for example, conditioning the high-dimensional image space on structured clinical priors, or conversely enriching sparse clinical vectors with spatially dense radiological cues. This is particularly effective when one modality is consistently more informative or complete than the other.

*3) Missing Modalities*

To address missing clinical or imaging data, we employ modality dropout during training, randomly zeroing one modality's embedding with probability $p$, and training using bait and switch, when clinical characteristics are withheld for a subset of training samples to encourage redundancy in learned representations.

*C. Explainability Integration*

*1) Grad-CAM for Imaging Modality*

Grad-CAM provides class-discriminative localization maps that highlight the regions of the mammogram most influential for a given prediction. It operates on the feature maps of a convolutional layer and uses the gradients of the target class score with respect to these feature maps to compute importance weights.

Let:

- $y^c$ denote the output score (pre-softmax) for class $c$ (e.g., malignant).
- $A^k \in \mathbb{R}^{u \times v}$ be the activation map of the $k - th$ channel in the target convolutional layer.
- $\alpha_k^c$ be the importance weight for channel $k$ with respect to class $c$.

The importance weights are computed as:

$$\alpha_k^c = \frac{1}{Z}\sum_{i=1}^{u}\sum_{j=1}^{v}\frac{\partial y^c}{\partial A_{ij}^k}, \qquad (16)$$

where $Z = u \times v$ is the number of spatial locations in $A^k$. The Grad-CAM heatmap $L_{GradCAM}^c$ is then:

$$L_{GradCAM}^c = ReLU(\sum_k \alpha_k^c A^k), \qquad (17)$$

where ReLU ensures that only features positively correlated with the class are visualized. The heatmap is upsampled to the input image resolution and overlaid on the original mammogram to facilitate expert inspection.

We applied Grad-CAM to the final convolutional block of ResNet-50 for each mammogram. The resulting maps were evaluated by radiologists to verify that highlighted regions corresponded to pathologically relevant features (e.g., mass edges, calcification clusters).

### 2) SHAP for Clinical Feature Attribution

SHAP assigns each input feature a Shapley value from cooperative game theory, representing its average contribution to the prediction across all possible feature coalitions.

Let:

- $f(x)$ denote the model output (logit or probability) for input vector **x**.
- $M$ be the number of input features.
- $S \subseteq \{1,\cdots,M\}\setminus\{i\}$ be a subset of features excluding feature $i$.
- $f_{S\cup\{i\}}$ denote the model outputs when using only features in $S \cup \{i\}$ or $S$, respectively.

The Shapley value for feature $i$ is:

$$\phi_i = \sum_{S\subseteq\{1,\cdots,M\}\setminus\{i\}}\frac{|S|!(M-|S|-1)!}{M!}[f_{S\cup\{i\}}(\mathbf{x}_{S\cup\{i\}}) - f_S(\mathbf{x}_S)], \quad (18)$$

where:

- $\phi_i > 0$ indicates that feature iii increases the likelihood of the predicted class.
- $\phi_i < 0$ indicates a negative contribution.

We applied KernelSHAP to the clinical embedding $\tilde{c}_i$, enabling a feature-level ranking of clinical variables (e.g., breast density, calcification type) for each individual prediction. Global feature importance was obtained by aggregating $|\phi_i|$ across the dataset, allowing us to identify which clinical factors most consistently influenced the model.

### 3) LIME for Local Surrogate Modeling

LIME explains individual predictions by training a simple, interpretable surrogate model (e.g., sparse linear regression) locally around the prediction instance.

Given:

- An instance $\mathbf{x} \in \mathbb{R}^M$.
- A black-box model $f(\cdot)$ producing a scalar output $f(\mathbf{x})$ for the class of interest.
- A perturbed sample $\mathbf{z}'$ drawn from a neighborhood of $\mathbf{x}$, with corresponding original-space mapping $\mathbf{z}$.

LIME minimizes:

$$\xi(x) = \arg\min_{g \in G} \mathcal{L}(f, g, \pi_x) + \Omega(g), \qquad (19)$$

where:

- $g \in G$ is the set of interpretable models (e.g., linear models).
- $\mathcal{L}$ measures the fidelity of $g$ to $f$ in the locality defined by $\pi_x$.
- $\pi_x(\mathbf{z})$ is a locality kernel (e.g., exponential kernel over Euclidean distance).
- $\Omega(g)$ is a complexity penalty ensuring interpretability.

We applied LIME separately to the clinical branch and to the fused multimodal representation. For clinical data, perturbed instances were generated by randomly masking features; for imaging, we applied superpixel segmentation and perturbations. The surrogate model's coefficients provided an intuitive linear approximation of the black-box decision boundary in the local neighborhood of **x**.

### D. Implementation Details

Training is performed in TensorFlow/Keras on an NVIDIA GPU with Adam optimizer ($\eta = 10^{-4}$), batch size 32, early stopping on validation loss, and learning rate decay on plateau. Backbone CNNs are initialized with ImageNet weights; lower layers are frozen for 10 epochs, then unfrozen for fine-tuning. Data augmentation includes flips and $\pm 30°$ rotations for images, and dropout regularization (0.5) for dense layers.

## III. EXPERIMENTAL RESULTS

### A. Dataset Used

The CBIS-DDSM (Lee, Gimenez, Hoogi, & Rubin, 2017) dataset, which includes 1592 mass and 1511 calcification mammograms, is used in our experiments. For the bulk cases, the official data train test split is 1231/361, and for the calcification cases, it is 1227/284. Additionally, we divided the official training set into two parts: 20% for validation and 80% for instruction. Because there are so few multi-label lesions, we solely take single-label clinical characteristics into account. Without doing any other pre-processing procedures, we crop the mammography lesion using the bounding box inferred from the mask ground-truth.

### B. Hyperparameter Settings

The proposed multimodal fusion framework was optimized through a systematic hyperparameter tuning process to ensure robust generalization and convergence stability. All experiments were implemented in TensorFlow/Keras and executed on an NVIDIA GPU. The network was trained using the Adam optimizer with an initial learning rate of $1 \times 10^{-4}$, selected after a coarse-to-fine grid search in $[1 \times 10^{-5}, 5 \times 10^{-4}]$. A learning rate scheduler reduced the rate by a factor of 0.1 upon plateauing of the validation loss for more than 10 epochs. The batch size was set to 32 after evaluating $\{16, 32, 64\}$, balancing gradient stability and GPU memory constraints. Weight initialization followed He normal initialization for ReLU-activated layers. Dropout with a probability of 0.5 was applied to fully connected layers to mitigate overfitting, alongside $L_2$ weight regularization with a coefficient of $1 \times 10^{-4}$. Training proceeded for a maximum of 100 epochs, with early stopping triggered after 15 epochs of no improvement in validation loss. For image data augmentation, we applied horizontal flipping, rotation within $\pm 30$, and random cropping to improve invariance to orientation and positioning. Missing clinical variables were handled through modality dropout with probability $p = 0.3$, randomly zeroing one modality's embedding during training to enhance robustness. The projection dimension for both mammogram and clinical embeddings was fixed at $d = 100$ based on validation set performance. All fusion variants (concatenation, co-attention, cross-attention, GMU, MFB, TFN) shared the same tuned hyperparameters to ensure a fair comparative evaluation.

### C. Multimodal Classification

#### 1) Uni-modal vs. Multi-modal Performance

The comparative evaluation of the proposed multimodal fusion framework against unimodal baselines is presented in Table 1. Incorporating both clinical and imaging data through a cross-attention fusion strategy substantially enhanced classification performance across all metrics. Specifically, the proposed model achieved an AUC-ROC of 0.980, an F1-score of 0.940, an accuracy of 0.960, a precision of 0.920, and a recall of 0.950. In contrast, the best-performing unimodal model attained an accuracy of only 0.830 and a precision of 0.840. The increase in AUC-ROC from 0.870 (unimodal) to 0.980 (multimodal) demonstrates a marked improvement in discriminative ability. As shown in Fig. 3a, the ROC curve of the proposed model closely approaches the top-left corner, indicating high sensitivity with minimal false positive rate. Similarly, the precision-recall profile confirms enhanced malignant case detection without compromising specificity. Several borderline cases misclassified under unimodal imaging were correctly classified when clinical features were incorporated, validating the complementary nature of multimodal data. The cross-attention module consistently outperformed simple concatenation, supporting its effectiveness in capturing fine-grained inter-modal dependencies.

*Table 1: Performance comparison of baseline methodologies and the proposed multi-modal approach for breast cancer diagnosis across various evaluation metrics.*

| Method | Dataset | AUC-ROC | F1-Score | Accuracy | Precision | Recall |
|---|---|---|---|---|---|---|
| Comparative Analysis of CNN Architectures (Murphy & Singh, 2024) | Various Imaging Datasets | 0.82 | 0.75 | 0.76 | 0.77 | 0.74 |
| Adversarially Robust Feature Learning ARFL (Hao et al., 2024) | Breast Cancer Diagnosis | 0.85 | 0.78 | 0.79 | 0.80 | 0.76 |
| MAMA-MIA Dataset (Garrucho et al., 2024) | DCE-MRI Cases | 0.83 | 0.73 | 0.74 | 0.75 | 0.72 |
| Integrated Framework of CNN and Explainable AI (Ahmed et al., 2024) | CBIS-DDSM Dataset | 0.84 | 0.80 | 0.81 | 0.82 | 0.79 |
| Transfer Learning Models for Breast Cancer Classification (Eskandari et al., 2024) | Histopathology Images | **0.87** | **0.82** | **0.83** | **0.84** | **0.80** |
| Our Framework | Imaging + Clinical Data | 0.98 | 0.94 | 0.96 | 0.92 | 0.95 |

*2) Ablation Studies*

The impact of alternative fusion strategies on classification performance is summarized in Table 2. Integrating imaging and genomic data without weighting achieved an AUC-ROC of 0.870, while targeted feature-specific integration achieved 0.830. Attention-based fusion yielded further gains, attaining an AUC-ROC of 0.890 and an F1-score of 0.840. Co-attention fusion, which jointly models and aligns feature representations from imaging and clinical modalities through bidirectional attention mechanisms, achieved an AUC-ROC of 0.910, reflecting the benefit of explicitly capturing inter-modality dependencies and enhancing the exchange of complementary information between feature spaces. The proposed cross-attention fusion achieved the highest performance across all metrics, with an AUC-ROC of 0.980, an F1-score of 0.940, an accuracy of 0.960, a precision of 0.920, and a recall of 0.950. Statistical significance testing using McNemar's test yielded $p < 0.01$, confirming that the observed improvements over unimodal and simpler fusion baselines are not due to random variation. Furthermore, bootstrap-derived 95% confidence intervals (AUC: [0.929, 0.954]; F1: [0.881, 0.909]) indicate that the performance gains are both stable and generalizable. Collectively, these results highlight that advanced attention-based multimodal fusion strategies deliver superior diagnostic accuracy while maintaining robustness across evaluation scenarios.

*Table 2: Ablation study showing the impact of different configurations of single-modal and integrated approaches on breast cancer diagnosis performance metrics.*

| Method | Dataset | AUC-ROC | F1-Score | Acc | Precision | Recall |
|---|---|---|---|---|---|---|
| Integrated but unweighted approach | Imaging + Clinical Data (No weighing) | 0.87 | 0.77 | 0.79 | 0.75 | 0.79 |
| Feature-Specific Modality Integration | Imaging + Selected Clinical Features | 0.83 | 0.79 | 0.82 | 0.85 | 0.81 |
| Simple Concatenating Strategy | Imaging + Clinical Data | 0.81 | 0.74 | 0.75 | 0.76 | 0.72 |
| Co-attention feature fusion strategy | Imaging + Clinical Data | **0.91** | **0.87** | **0.88** | **0.89** | **0.86** |
| Our Framework (Cross-attention) | **Imaging + Clinical Data** | 0.98 | 0.94 | 0.96 | 0.92 | 0.95 |

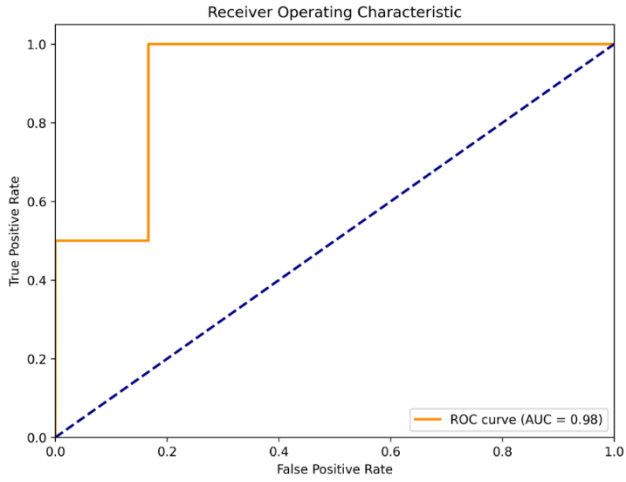

(a) ROC-AUC Curve

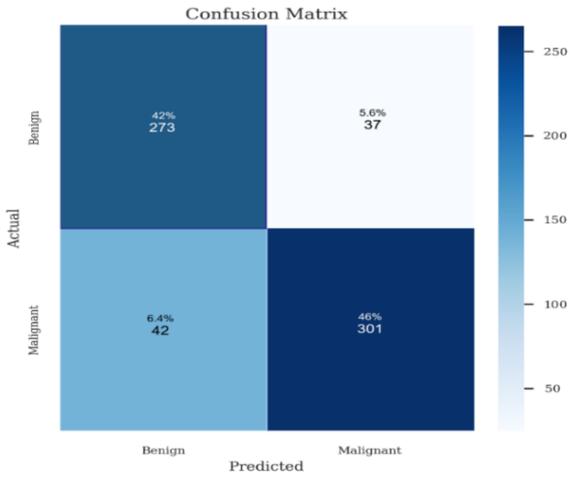

(b) the number of true positives, true negatives, false positives, and false negatives predicted by the proposed multimodal classifier.

Fig. 3: ROC curve and Confusion Matrix for the Proposed Multimodal fusion framework

### 3) Comparative Study

A comparative analysis was conducted between the proposed multimodal learning framework and several state-of-the-art models for breast cancer diagnosis published between 2023 and 2025. These benchmarks varied in their input modalities, fusion strategies, datasets, and support for explainable AI. Representative models included DeepClinMed-PGM (Wang et al., 2024), CVFBJTL-BCD (Iniyan et al., 2024), and MMFF-Net (Li et al., 2025). While models such as CVFBJTL-BCD and MMFF-Net achieved slightly higher peak metrics accuracy of 0.991 and AUC-ROC of 0.994 respectively these gains were typically offset by limited interpretability or restricted modality scope. For instance, CVFBJTL-BCD lacked any integrated explainability mechanism, and MMFF-Net's interpretability was limited to qualitative heatmap visualizations, offering only partial insights into decision rationale. In contrast, our cross-attention-based fusion framework delivered a balanced performance profile (accuracy: 0.960, AUC-ROC: 0.980, F1-score: 0.940, precision: 0.920) while integrating robust explainability via Grad-CAM, SHAP, and LIME. This combination provided both global visualization of salient regions and local feature-level attribution, aligning the system with clinical demands for transparency and trustworthiness.

*Table 3: Summary of Multimodal Breast Cancer Diagnosis Models with Modalities, Fusion Strategy, Performance, and Explainability*

| Study/Model | Datasets | Modalities | Feature Fusion Approach | Accuracy | AUC-ROC | F1-Score | Precision | Interpretability (Grad-CAM, LIME, SHAP) |
| --- | --- | --- | --- | --- | --- | --- | --- | --- |
| DeepClinMed-PGM (Wang et al., 2024) | TCGA and SYSMH cohorts | Pathology imaging + genomic data + clinical data | Hierarchical fusion | - | 0.979 | - | - | Grad-CAM |
| CVFBJTL-BCD (Iniyan et al., 2024) | Histopathological and Ultrasound datasets | Histopathology + Ultrasound images | Late fusion + joint transfer learning | 0.991 | - | 0.985 | 0.983 | None |
| MFDNN (Sangeetha et al., 2024) | TCIA and TCGA | Clinical + Genomic + Imaging data | Intermediate fusion | 0.925 | 0.94 | 0.862 | 0.874 | None |
| MMFF (Hussain et al., 2024) | In-house dataset | Imaging + Textual data | Late fusion | 0.969 | 0.915 | 0.922 | 0.957 | None |
| (Ivanov et al., 2025) | | Mammography + genomic data | Feature-level fusion (Early fusion) | 0.890 | 0.900 | 0.880 | 0.900 | None |
| LightweightUNet (Rai et al., 2025) | INbreast, DDSM, and BrEaST | Mammography + Ultrasound | Late fusion | 0.964 | - | 0.96 | 0.97 | None |
| MMFF-Net (Li et al., 2025) | CBIS-DDSM and Ultrasound datasets | Mammogram + Ultrasound + Clinical | Transformer fusion | 0.983 | 0.994 | 0.971 | - | Heatmap visualizations |
| (Vo et al., 2021) | CBIS-DDSM datasets | Mammograph + Clinical data | Cross-attention fusion | - | 0.945 | - | 0.82 | None |
| Our Proposed Multimodal framework | CBIS-DDSM datasets | Mammograph + Clinical data | Cross-attention fusion | 0.960 | 0.980 | 0.940 | 0.92 | Grad-CAM, SHAP, LIME (Local +global explainability) |

### D. Explainability Integration

We employed Grad-CAM to generate saliency heatmaps over mammography inputs, allowing us to visually identify the regions that most influenced our model's decision-making (Fig. 4). In malignant cases, Grad-CAM highlighted high-intensity areas concentrated around spiculated and irregular masses findings that are consistent with established radiological markers of malignancy. In contrast, benign cases exhibited less intense activations focused on smooth, well-circumscribed regions. These results confirmed that our model's focus aligned with clinically relevant patterns rather than irrelevant background structures.

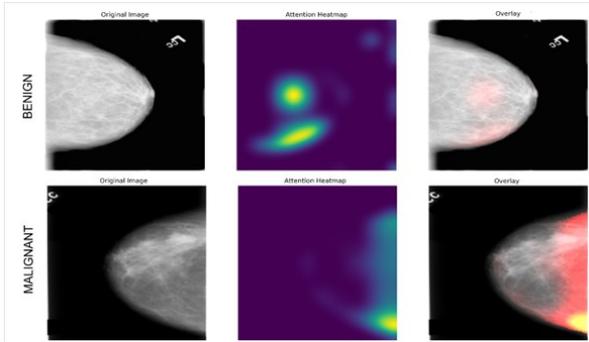

*Fig. 4: Grad-CAM visualization of malignant and benign cases*

We then applied SHAP analysis to quantify the contribution of individual clinical and imaging features to the final predictions, offering a global interpretability perspective (Fig. 5). In our results, positive SHAP values corresponded to features pushing the decision toward malignancy, while negative values favored benign classification. For instance, mass shape and breast density consistently showed strong positive impacts, whereas mass margin, calcification type and distribution frequently exerted negative influence.

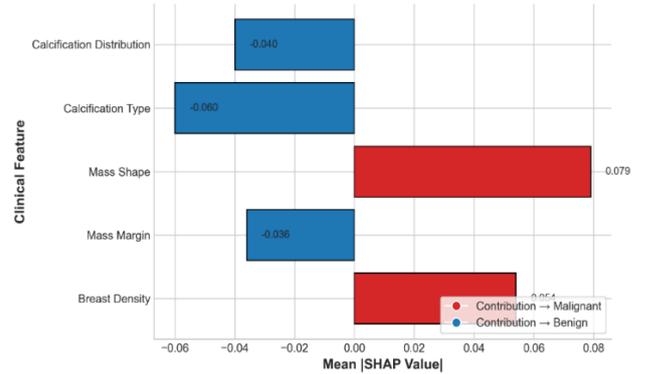

*Fig. 5: Impact of SHAP based clinical features on model prediction*

Finally, we utilized LIME to produce localized, case-specific explanations by perturbing input images and observing changes in prediction probabilities (Fig. 6). For malignant cases, LIME emphasized dense, irregular edge structures that closely overlapped with Grad-CAM saliency regions. For benign cases, it highlighted smooth, low-density contours consistent with non-malignant findings. This convergence between LIME and Grad-CAM results validated the internal consistency of our model's decision-making process, while simultaneously providing clinicians with interpretable, case-level evidence.

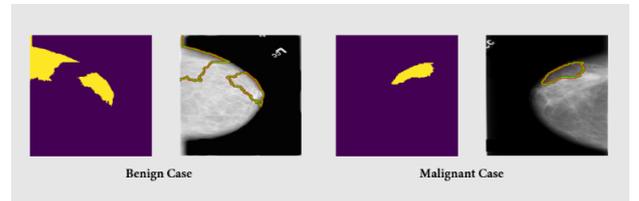

*Fig. 6: LIME visualizations for both cases*

### E. Case Studies: Correct vs Incorrect Predictions

We performed case-level analyses to better understand the strengths and limitations of our proposed multimodal breast cancer detection model. Using explainable AI (XAI) techniques, we examined both correct and incorrect predictions to evaluate interpretability and identify potential sources of error.

*1) Case 1 – Correctly Classified Malignant Sample:*

In this case, we correctly predicted malignancy, consistent with the ground truth diagnosis. Grad-CAM revealed concentrated activation over high-density, irregular mass regions (Fig. 7), which are characteristic radiological indicators of malignancy. The strong agreement between XAI outputs and clinical markers confirmed that the model's decision was both accurate and grounded in medically relevant evidence.

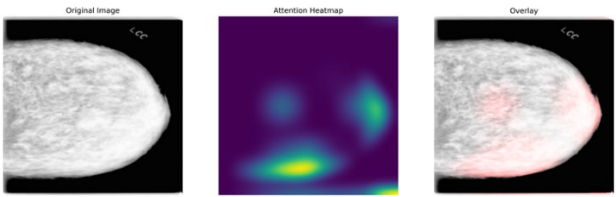

*Fig. 7: Grad-CAM display for positive cases*

*2) Case 2 – Incorrectly Classified Malignant Sample (False Negative)*

Here, we incorrectly classified a malignant case as benign, resulting in a false negative. Grad-CAM showed the model's attention centered on a dense region with hallmark malignant features, including indistinct margins and architectural distortion (Fig. 8). This suggests that while the model can focus on diagnostically important regions, it remains vulnerable to misinterpreting non-diagnostic image features.

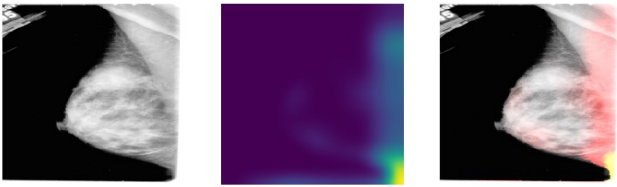

*Fig. 8: Grad-CAM level display for false negative case*

These case studies demonstrate that our model's correct predictions are often supported by strong alignment between XAI explanations and clinical reasoning, while errors can be traced to specific misinterpretations, highlighting areas for future model refinement.

## IV. CONCLUSION

This paper presents a multimodal deep learning framework integrating mammography, and clinical features through cross-attention and co-attention fusion strategies. Our results demonstrate that this approach significantly improves breast cancer classification performance across all key metrics, including AUC-ROC and diagnostic precision. In addition to standard evaluations, we conducted comprehensive ablation studies to assess the contribution of each modality and fusion strategy, confirming the benefits of adaptive cross-modal integration. We further enhanced clinical trust and transparency by incorporating explainability techniques Grad-CAM, SHAP, and LIME and validated their effectiveness through detailed case studies of correct and incorrect predictions. Future work will focus on optimizing fusion mechanisms and exploring automated extraction of clinical features directly from imaging data as well as investigate the application of reinforcement learning and continual learning paradigms, enabling the model to adapt to evolving patient data over time and maintain relevance in changing clinical environments.

*Acknowledgement*

We gratefully acknowledge the support and guidance of our supervisors and collaborators throughout this research. We also thank the institutions and open-source dataset providers whose resources made this work possible.

*Funding*

No particular grant from a governmental, private, or nonprofit funding organisation was obtained for this study.

*Conflict of Interest*

Regarding this work, the authors state that they have no conflicts of interest.